\documentclass[%
 reprint,
 superscriptaddress,
 amsmath,amssymb,
 aps,
]{revtex4-1}

\usepackage{graphicx}
\usepackage{dcolumn}
\usepackage{bm}
\usepackage{color}
\usepackage{xcolor}
\usepackage{soul}
\usepackage{subcaption}
\usepackage[utf8]{inputenc}

\begin{document}


\title{Superheated solid state induced by a single collision event} 

\author{Claudia Loyola}
 \email{claudia.loyola@unab.cl}
\affiliation{Departamento de F\'isica, Facultad de Ciencias Exactas, Universidad Andr\'es Bello, Sazi\'e 2212, Santiago, Chile.}

\author{Sergio Davis}
 \affiliation{Comisi\'on Chilena de Energ\'ia Nuclear, Nueva Bilbao 12500, Santiago, Chile.}
 \affiliation{Departamento de F\'isica, Facultad de Ciencias Exactas, Universidad Andr\'es Bello, Sazi\'e 2212, Santiago, Chile.}
 \email{sergio.davis@cchen.cl}

\author{Joaqu\'in Peralta}
 \email{joaquin.peralta@unab.cl}
\affiliation{Departamento de F\'isica, Facultad de Ciencias Exactas, Universidad Andr\'es Bello, Sazi\'e 2212, Santiago, Chile.}

\date{\today}

\begin{abstract}
High-energy incident particles from both pulsed and continuous radiation sources can induce significant damage to the structure of a material by creating vacancy-interstitial
pairs and other more complex defects, and this leads typically to localized melting. In this work, we present evidence via molecular dynamics simulations of a superheated solid state
in BCC tungsten induced by single PKA events of $\sim$ 1.5 keV of energy. Despite the striking difference between this highly inhomogeneous energy injection and homogeneous melting, the lifetime of the obtained superheated solid state, reaching up to 200 ps, is described through a waiting time distribution in agreement with previous studies on superheating in the Z-method methodology.
\end{abstract}

\maketitle

\section{\label{sec:Introduction}Introduction}

The study of damage to structural materials produced by irradiation is a long-standing research topic~\cite{Nordlund2018}. Radiation damage in tungsten (W), in particular, has attracted
considerable attention as W appears to be one of the top candidates for the plasma-facing walls of nuclear fusion reactors~\cite{Philipps2011,InestrosaIzurieta2015,Valles2017,Fellman2019}.
Under high enough doses of radiation, it is expected that the material can, after melting locally, either recrystallize or remain in an amorphous state. This nonequilibrium
melting process might provide some insight into the process of formation of vacancy-interstitial pairs.

Atomistic computer simulation of radiation damage often proceeds using an approximation called \emph{primary knock-on atom} (PKA)~\cite{Fikar2009,Sand2013,Nordlund2019}. In this method, a
single atom at random is forcefully displaced from its equilibrium lattice position, representing the collision effect with a high-energy incident particle. After being displaced
from its initial position, the PKA can induce subsequent displacements of other atoms given enough initial energy.

On the other hand, nonequilibrium melting can also be studied as a perfectly homogeneous process, where a total amount of kinetic energy is given to a perfectly ideal crystal.
A well-known computational example of this methodology used to determine the melting curves of materials is the so-called Z-method~\cite{Belonoshko2006b}, in which a metastable
(superheated) solid state can be obtained with considerable lifetimes~\cite{Alfe2011,Davis2018d}. In this case, it is known that the formation of vacancy-interstitial pairs and
their coordinated movement play an essential role in the stability of the superheated solid state, as shown in several previous works~\cite{Bai2008,Davis2011c,Zhang2013a}.

In this work, we study the kinetics and microscopic mechanism of melting from PKA events, compared to the kinetics of the homogeneous melting. We report the
counterintuitive observation that a superheated, metastable state in BCC W can be produced by a single, high energy collision, in a reproducible manner.

The rest of the paper is organized as follows. Section~\ref{sec:Computational} describes the molecular dynamics and statistical analysis methodologies used, while
Section~\ref{sec:Results} presents the main results and their interpretation. Finally, we close with some concluding remarks in Section~\ref{sec:Conclusions}.

\section{\label{sec:Computational}Computational Procedure}

Computer simulations were performed using the LAMMPS software package~\cite{Plimpton1995Mar}. A bulk structure of BCC W of $N=432$ atoms was considered with a lattice parameter
of $a=3.1803$~\AA, and periodic boundary conditions on its three axes $x$, $y$, and $z$. All simulations were performed at constant energy and constant volume (i.e., in the microcanonical
or NVE ensemble) with an integration time-step for the velocity-Verlet algorithm of $\Delta t$ = 0.1 fs, for a total duration of 200 ps.

The initial background temperature of the system, before the PKA event, was set to $T_{\text{BG}}$=0.001 K. The interatomic potential chosen to describe W under the effect of high-temperatures and its response to
radiation is the SNAP interatomic potential for W by Wood \emph{et al}~\cite{Wood2017Feb}. This type of interatomic potentials, based on machine-learning~\cite{Bartok2010Apr}, uses a
set of local descriptors to calculate the per-atom energy function. A complete description of SNAP potentials can be found in the work of Thompson \emph{et al}~\cite{Thompson2015Mar}. In
order to generate a reasonable statistical sample, we performed a set of 3000 independent simulations at the same PKA energy, as described in Section~\ref{sec:Results}.

\begin{figure}
\centering
\includegraphics[width=.45\textwidth]{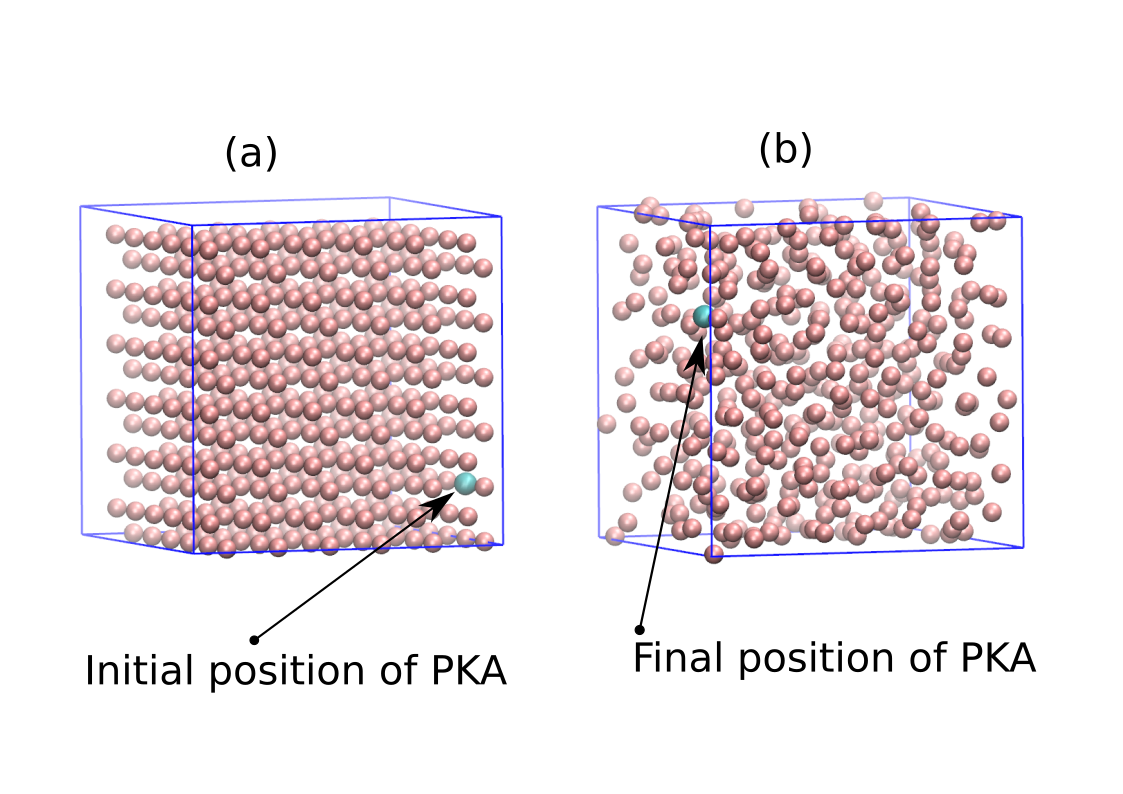}
\caption{(a) The initial configuration of the BCC W crystal of 432 atoms is observed. The blue, slightly larger atom corresponds to the PKA, which is set to higher energy at the initial
step of simulation. (b) Configuration following the PKA displacement, after 200 ps of simulation in the system.}
\label{fig:W-initial}
\end{figure}

Fig.~\ref{fig:W-initial} shows the initial structure of W with a random atom selected as the PKA, which receives the total energy at $t$=0. In the following we present the main
results obtained from the different simulations, summarized in the behavior of melting waiting times, structural properties, kinetics of vacancy formation, and
common neighbor analysis.

\section{\label{sec:Results}Results and Discussion}

\begin{figure}
\centering
\includegraphics[width=0.5\textwidth]{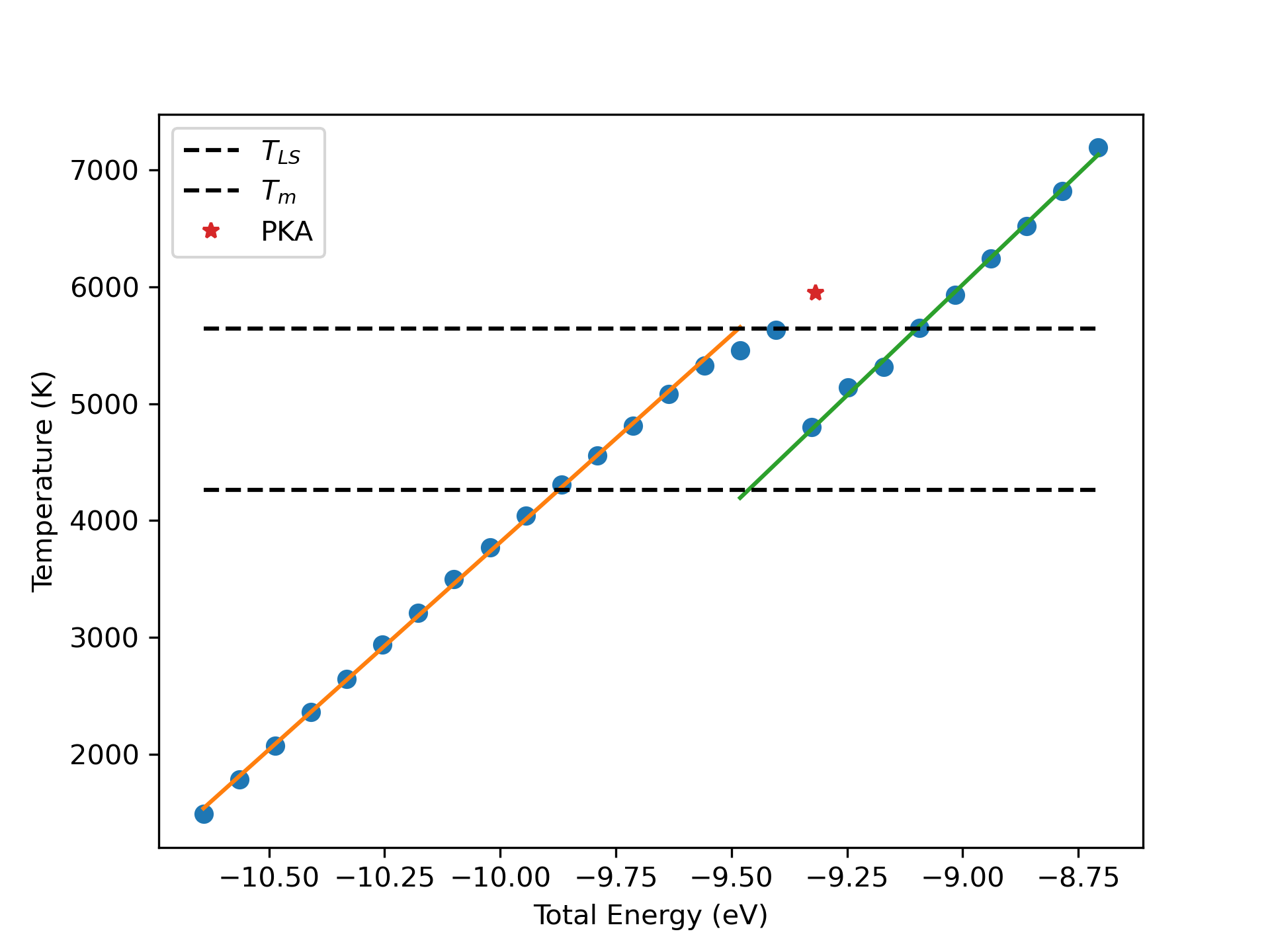}
\caption{Caloric curve $T(E)$ at constant volume. The circles represent equilibrium Z-method simulations, and the star indicates the nonequilibrium conditions
induced by the PKA event. Straight lines represent, from left to right, the solid and liquid branches.}
\label{fig:ZCurve}
\end{figure}

The constant-volume caloric curve (Z-curve), using the Z-method~\cite{Belonoshko2006b}, is presented in Fig.~\ref{fig:ZCurve} for the W system, showing the melting transition at
$T_m \sim $4700 K and the limit of superheating $T_{LS} \sim$6000 K. In this method, some kinetic energy is given in a homogeneous manner to the ideal crystal, which therefore remains
for the rest of the simulation at a total energy
\begin{equation}
E = \Phi_0 + \frac{3N}{2}k_B T_0,
\end{equation}
where $\Phi_0$ is the potential energy of the ideal crystal. In the case of PKA simulations such as the ones described in the rest of the paper, the total energy
in a simulation is instead given by
\begin{equation}
E = \Phi_0 + \frac{3N}{2}k_B T_{\text{BG}} + E_{\text{PKA}} \approx \Phi_0 + E_{\text{PKA}}.
\end{equation}

We considered an initial PKA energy $E_{\text{PKA}}$=1.478 keV, choosing the
displaced atom (as PKA) randomly, so that after receiving the energy from the PKA event the crystal structure reaches a total energy slightly above the limit of superheating, in this
case, $E=-9.318$ eV.

This energy of 1.478 keV corresponds to a value on the lower-end of radiation damage collision energies considered on other works~\cite{Sand2016}, which range from tens of eV to hundreds of keV. Despite the low collision energies, a spontaneous melting phenomenon is observed with the presence of
superheating of the crystalline structure. Our results clearly show that a considerable change in a reduced zone of the crystalline structure is not enough to
melt the whole crystal, and the structure remains as a critical superheated solid for some time.

\begin{figure} 
\centering
\includegraphics[width=0.5\textwidth]{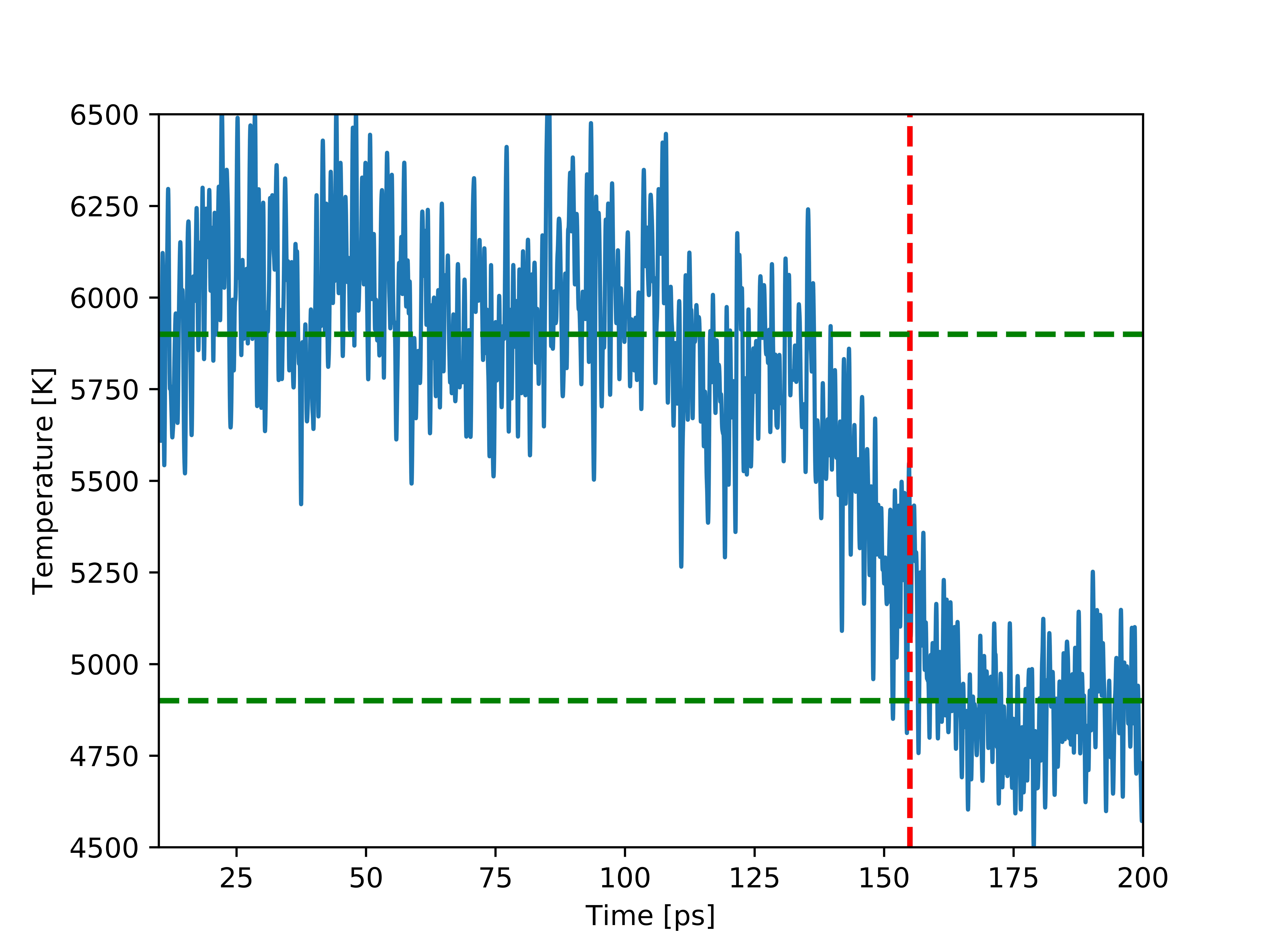}
\caption{Instantaneous temperature as a function of time for a single realization of the PKA at 1.478 keV. It can be seen that the metastable, superheated
solid lasts for up to 155 ps.}
\label{fig:TempTime}
\end{figure}

Fig.~\ref{fig:TempTime} shows the time evolution of the instantaneous temperature for the W crystal after the PKA event, for a single realization at 1.478 keV.
We can see that the collapse of the crystal occurs at approx. 155 ps. This defines two phases, a metastable phase at high temperature initiated by the PKA event, and a second,
liquid phase which occurs closely after 200 ps.

\begin{figure} 
\centering
\includegraphics[width=0.5\textwidth]{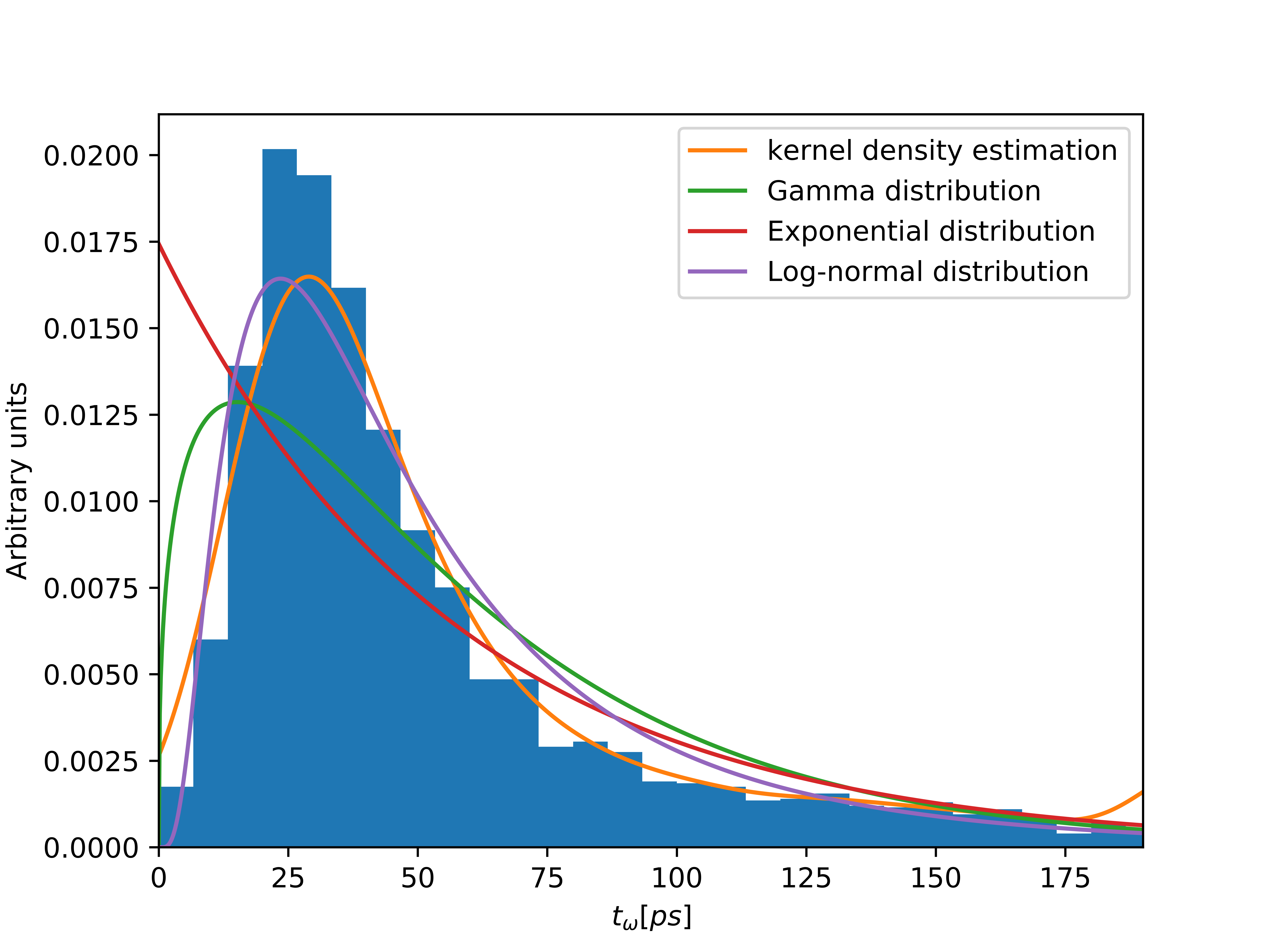}
\caption{Histogram of waiting times until spontaneous melting, for PKA events at 1.478 keV, together with exponential, log-normal and gamma model fits and a reference model
obtained using the kernel density estimation (KDE) method.}
\label{fig:Histogram1}
\end{figure}

\begin{figure} 
	\centering
	\includegraphics[width=0.5\textwidth]{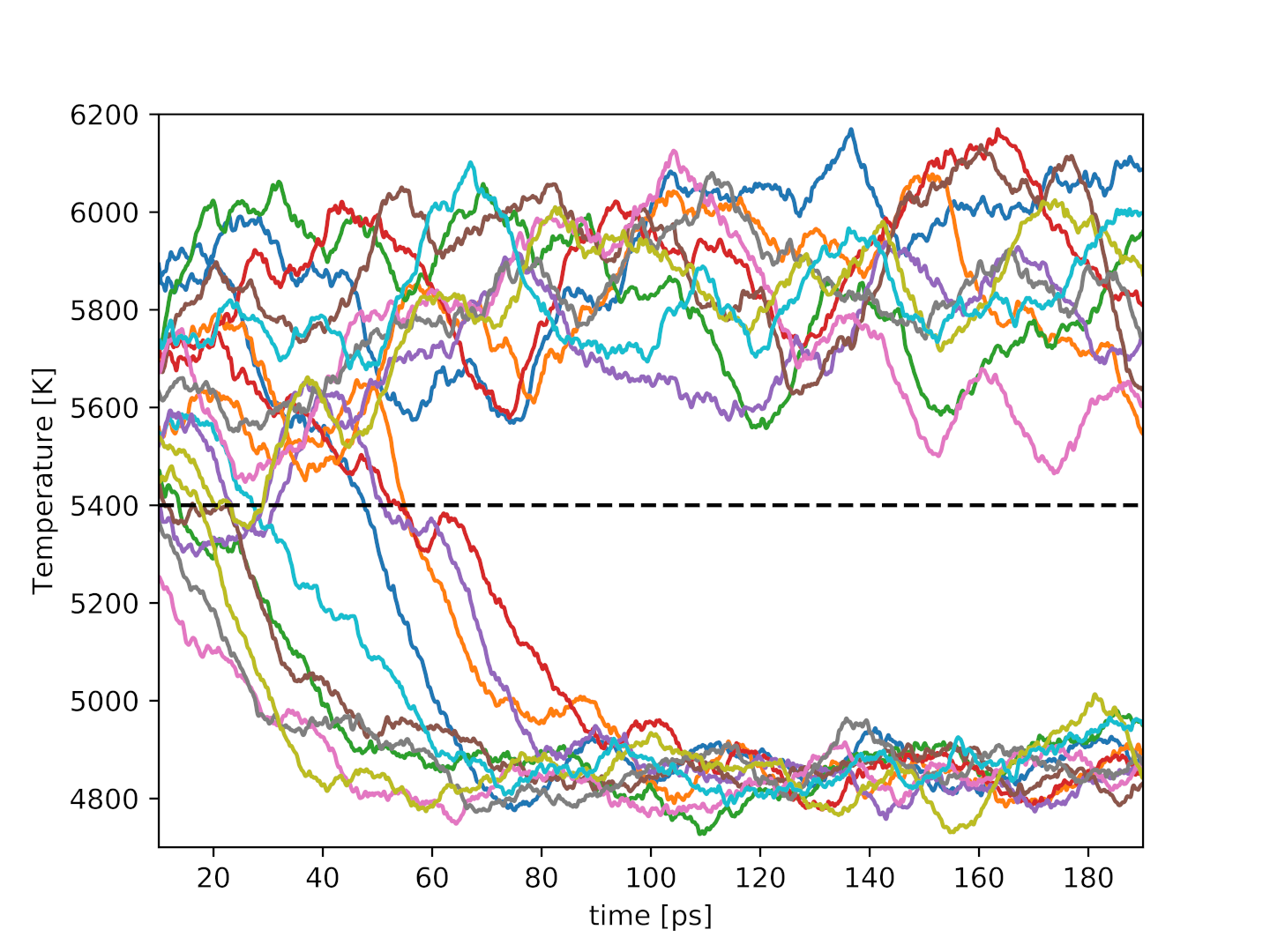}
	\caption{Simulation temperature for 20 different selected cases. There is 10 cases corresponding to the melting phenomena in which temperature drops to $\sim4700$ K, and the other 10 cases remains in a overheating state during the whole simulation, 200 ps.}
	\label{fig:Temps}
\end{figure}

The lifetime of the metastable phase (waiting times until the collapse of the crystal) is observed in Fig.~\ref{fig:Histogram1} from all 3000 MD simulations. The frequencies of the observed lifetimes strongly resemble those observed in the homogeneous superheated state~\cite{Davis2018d}, and theoretical models based on continuous-time
random walks~\cite{OlguinArias2019b}. Additionally a selected group of simulations of 10 superheated and 10 melting states were considered for structural analysis
of both phases. Figure \ref{fig:Temps}, shows the temperature during the whole simulation time for these cases. As we can observe, 10 of them drop before 100 ps to a
temperature $\sim4700$ K, which corresponds to the melting temperature of W for this system size and interatomic potential. On the other hand, the
rest of the selected cases preserve a critical superheated state which remains as solid at least for the simulation time of 200 ps. The black dashed line represents a
temperature in between the superheating limit and the melting temperature of the system.

\begin{figure} 
	\centering
	\includegraphics[width=0.5\textwidth]{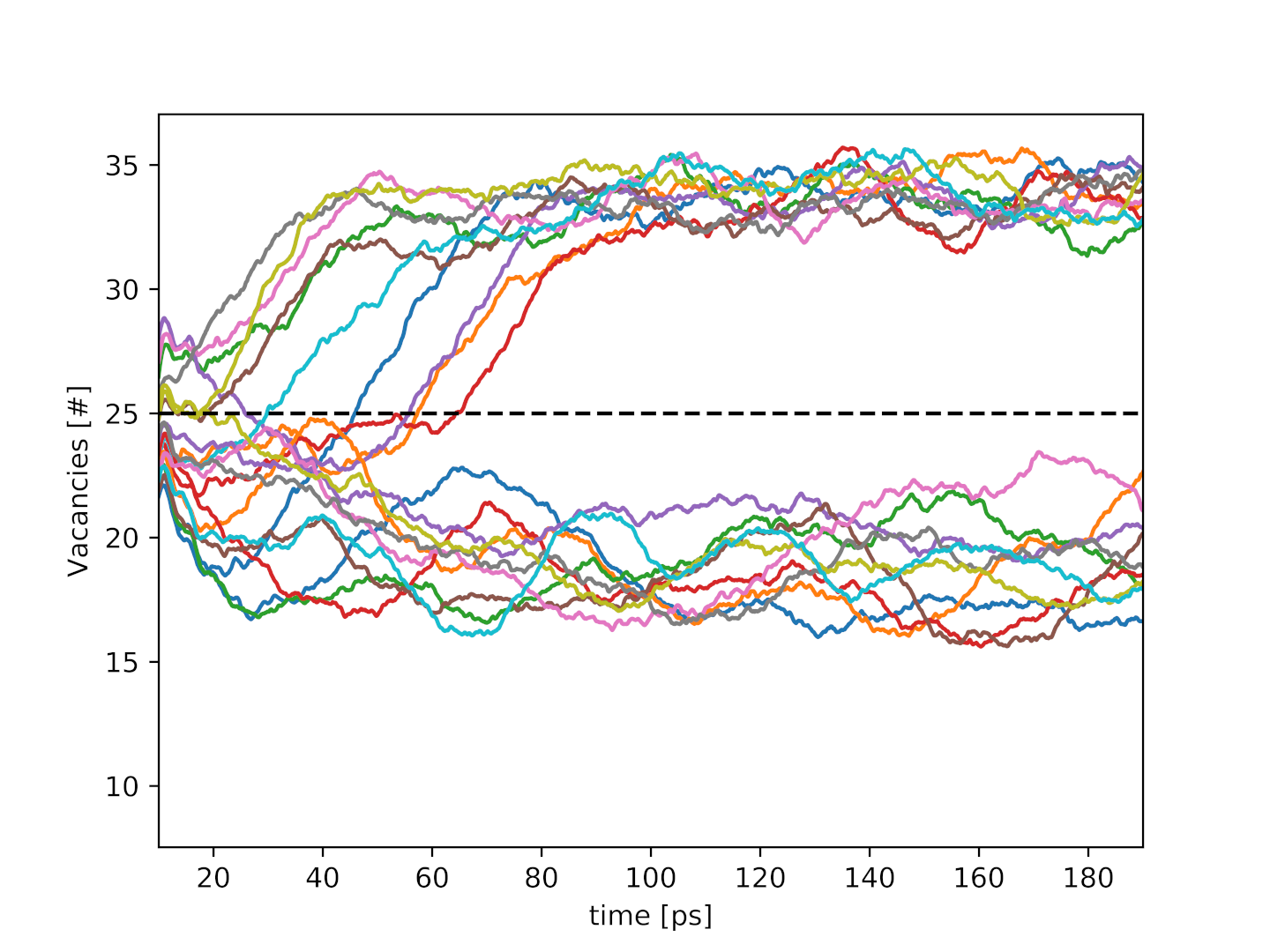}
	\caption{Different simulations with melting process occurring before 100 ps, and with overheating lasting by almost the full simulation time of 200 ps. Two branches are observed, characterizing the presence of thermal vacancies in both systems, however for values above the 25 vacancies melting is produced.}
	\label{fig:Vacancies}
\end{figure}

The formation of vacancy-interstitial pairs is also known to induce long-range correlations in the superheated solid~\cite{OlguinArias2019}. In order to detect correlated movement
in our metastable state we computed the kinetic evolution of thermal vacancies, using an improved GPU version~\cite{Peralta2015} of the SearchFill algorithm~\cite{Davis2011b},
which has already been tested on thermal vacancies below the melting point~\cite{Pozo2015}. Time-dependent behavior of the number of vacancies for a limited set of
20 samples, 10 for superheated and 10 for melting state, are presented in the Figure~\ref{fig:Vacancies}. Vacancies were evaluated every 0.04 ps, using a overlap
vacancy of $v_{ovp} = 0.54$, which was determined by the BCC crystalline structure at $\sim 4000$ K. Two main branches are observed in the vacancies, which are related to both metastable and
liquid state. When the number of vacancies increases above 25, the melting process is triggered and becomes irreversible, producing a liquid state in the sample. On the other
hand, when this number of vacancies remains below 25, a critical superheated state is observed during the whole simulation. The curves were smoothed using a convolution procedure
in order to remove statistical noise in the plot.

Structural properties of both metastable, and liquid phase are described by the concentration of common neighbor analysis (CNA). The CNA is a method for analyzing
structures by a decomposition of the pair-distribution function according to the local environment of the bonded pairs~\cite{Loyola2010}. The CNA is represented by diagrams that
are classified by a set of four indexes. The first index $i$ has values 1 or 2 indicating that a pair of atoms a and b are nearest neighbors $(i = 1)$ or not $(i = 2)$; the second index
$j$, indicates the number of neighbors common to both atoms ($a$ and $b$); the third index $k$ is the number of bonds between the common neighbors; and the last index $l$ is the length
of the longest continuous chain formed by those $k$ bonds. From the common neighbor analysis, in the ideal BCC structure, two principal parameters are highlighted, the $jkl$ triplets 444
and 555, both with $i=1$. In the perfect crystal a concentration of 42.857\% (3/7) of the pairs are of 444 type and the remaining 57.143\% (4/7) of the pairs are of 666 type.

\begin{figure*}[h!]
    \centering
    \begin{subfigure}[t]{0.45\textwidth}
        \centering
        \includegraphics[width=0.9\textwidth]{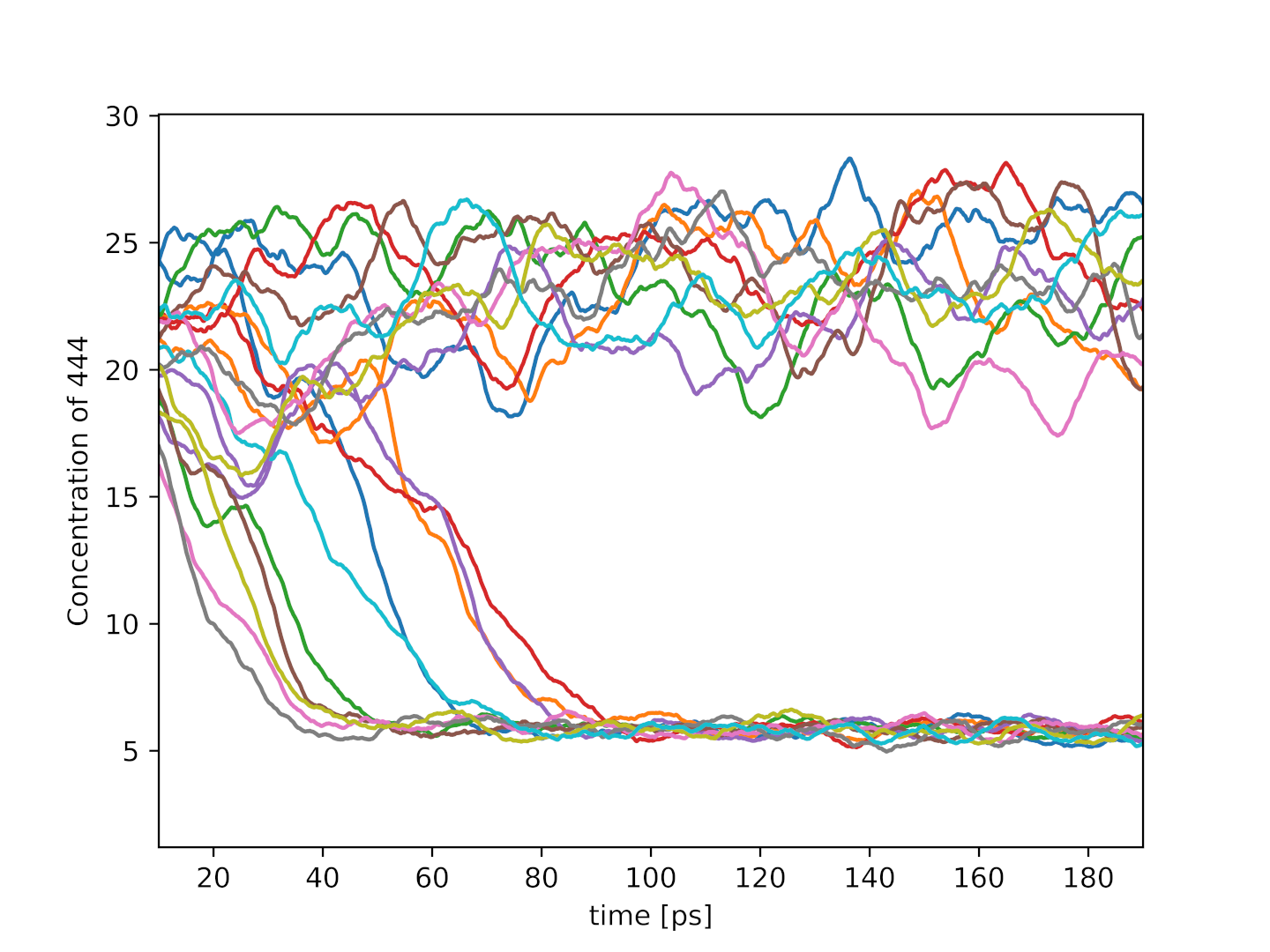}
        \caption{444}
    \end{subfigure}%
    ~
    \begin{subfigure}[t]{0.45\textwidth}
        \centering
        \includegraphics[width=0.9\textwidth]{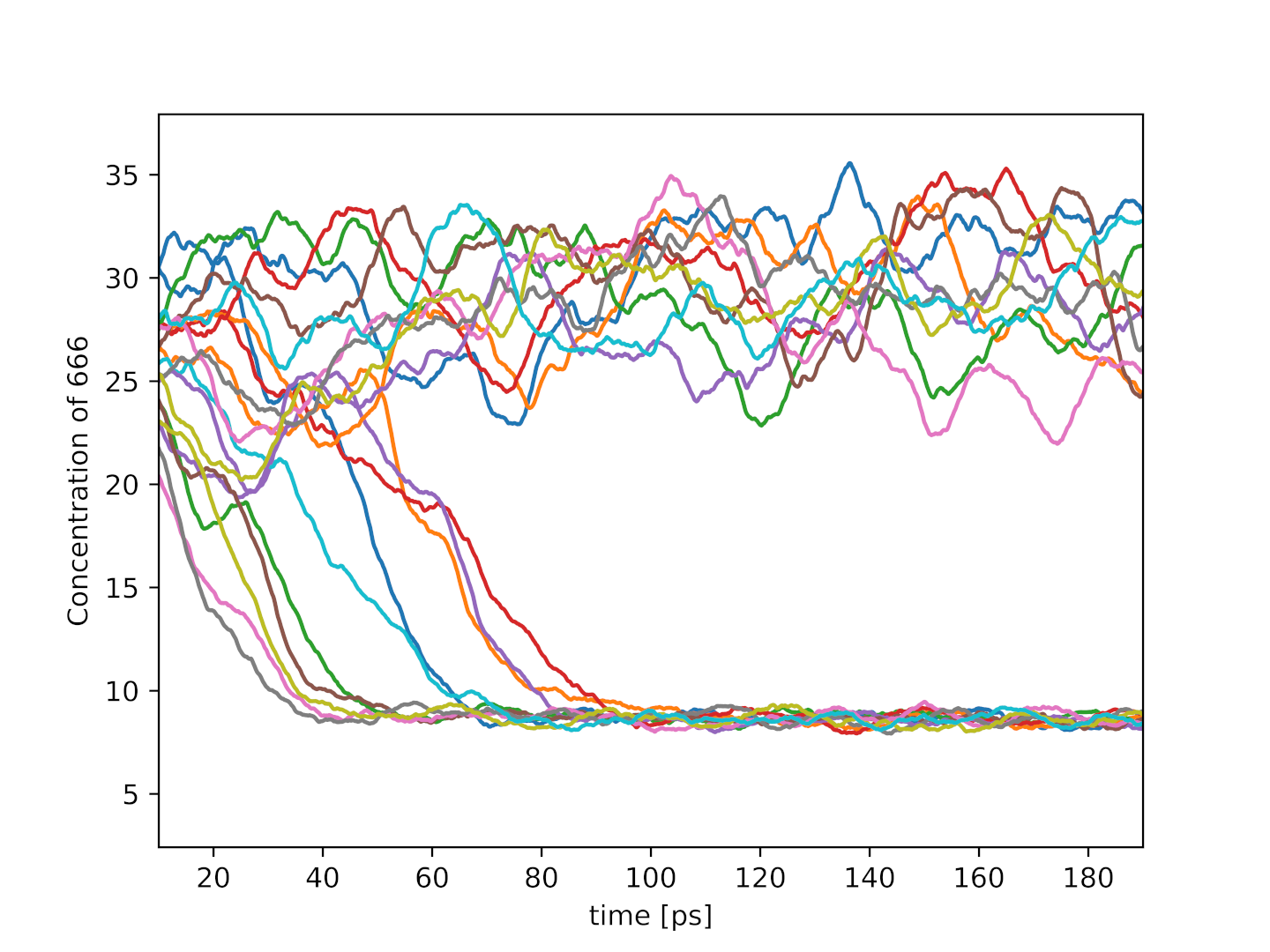}
        \caption{666}
    \end{subfigure}
    \caption{Configurations triplets related to the presence of a BCC crystal structure. Concentration of both triples, 444 and 666, drops to values of 5\% and 10\% respectively, when melting is reached.}
    \label{fig:CNA-BCC}
\end{figure*}

Figure~\ref{fig:CNA-BCC} shows the main CNA triplets considered in the BCC crystal for the 20 selected simulations, as described previously. The concentration of 444 and 666 triplets
drops rapidly, as expected, from their initial values in the ideal bcc structure around $\sim 25$\% and $\sim 30$\%, respectively, when the crystalline structure collapses and the
liquid state is formed. In this case the percentages drop to values close to $\sim 5$\% and $\sim 8$\% for the triplets 444 and 666, a clear indication of the complete conversion to
the liquid phase.

\begin{figure*}[h!]
    \centering
    \begin{subfigure}[t]{0.45\textwidth}
        \centering
        \includegraphics[width=0.9\textwidth]{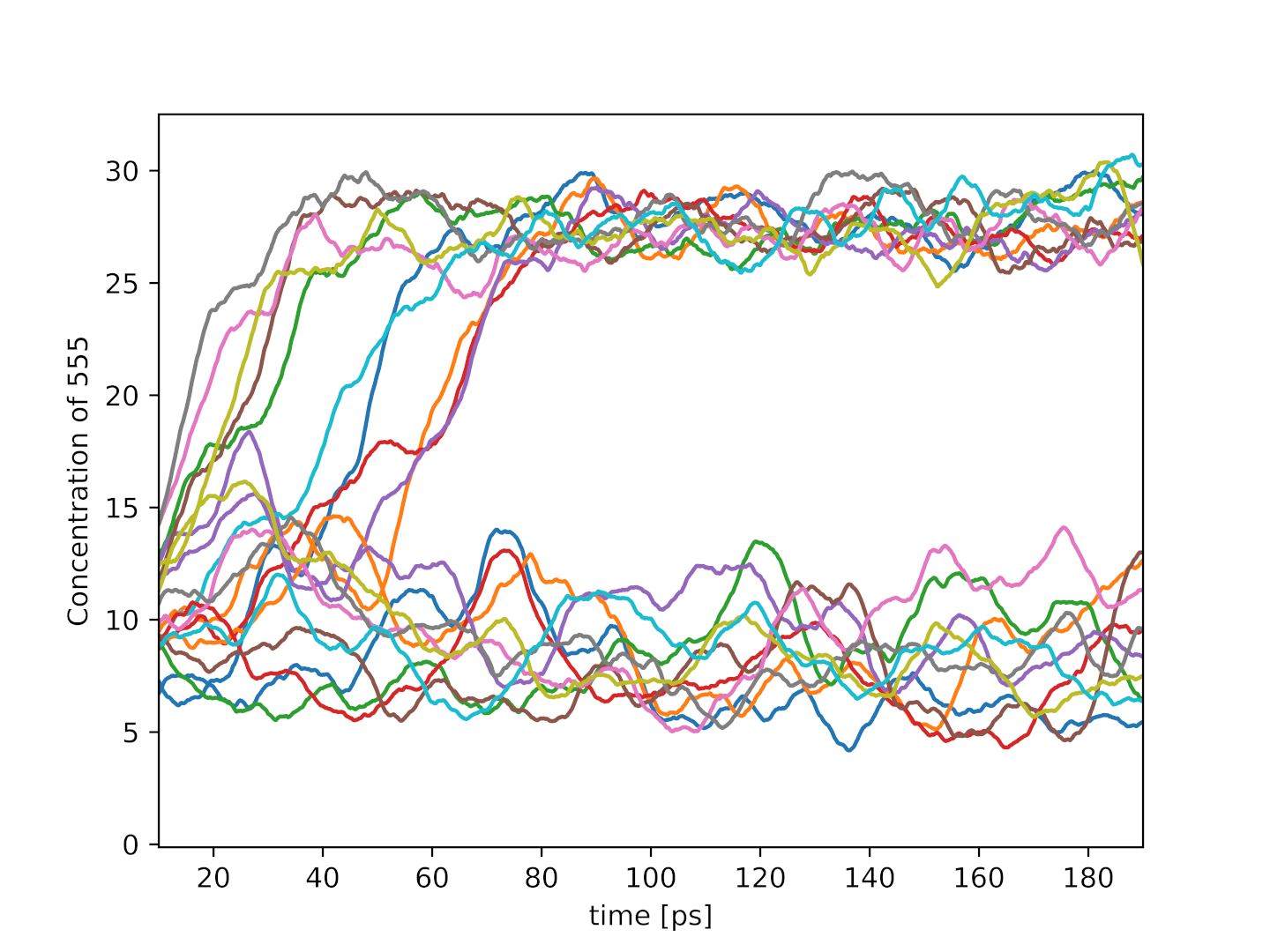}
        \caption{555}
    \end{subfigure}%
    ~
    \begin{subfigure}[t]{0.45\textwidth}
        \centering
        \includegraphics[width=0.9\textwidth]{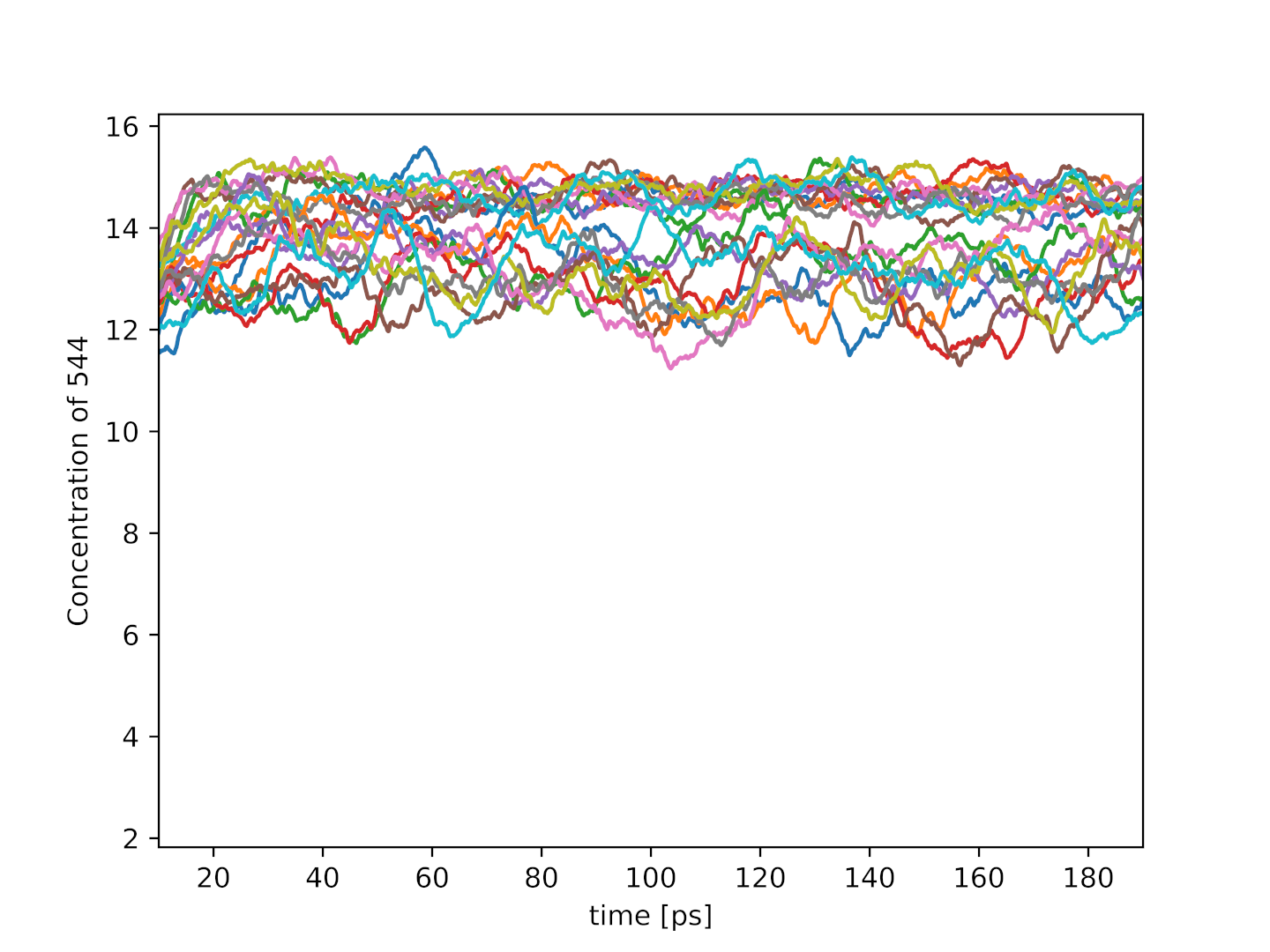}
        \caption{544}
    \end{subfigure}
      ~
    \begin{subfigure}[t]{0.45\textwidth}
        \centering
        \includegraphics[width=0.9\textwidth]{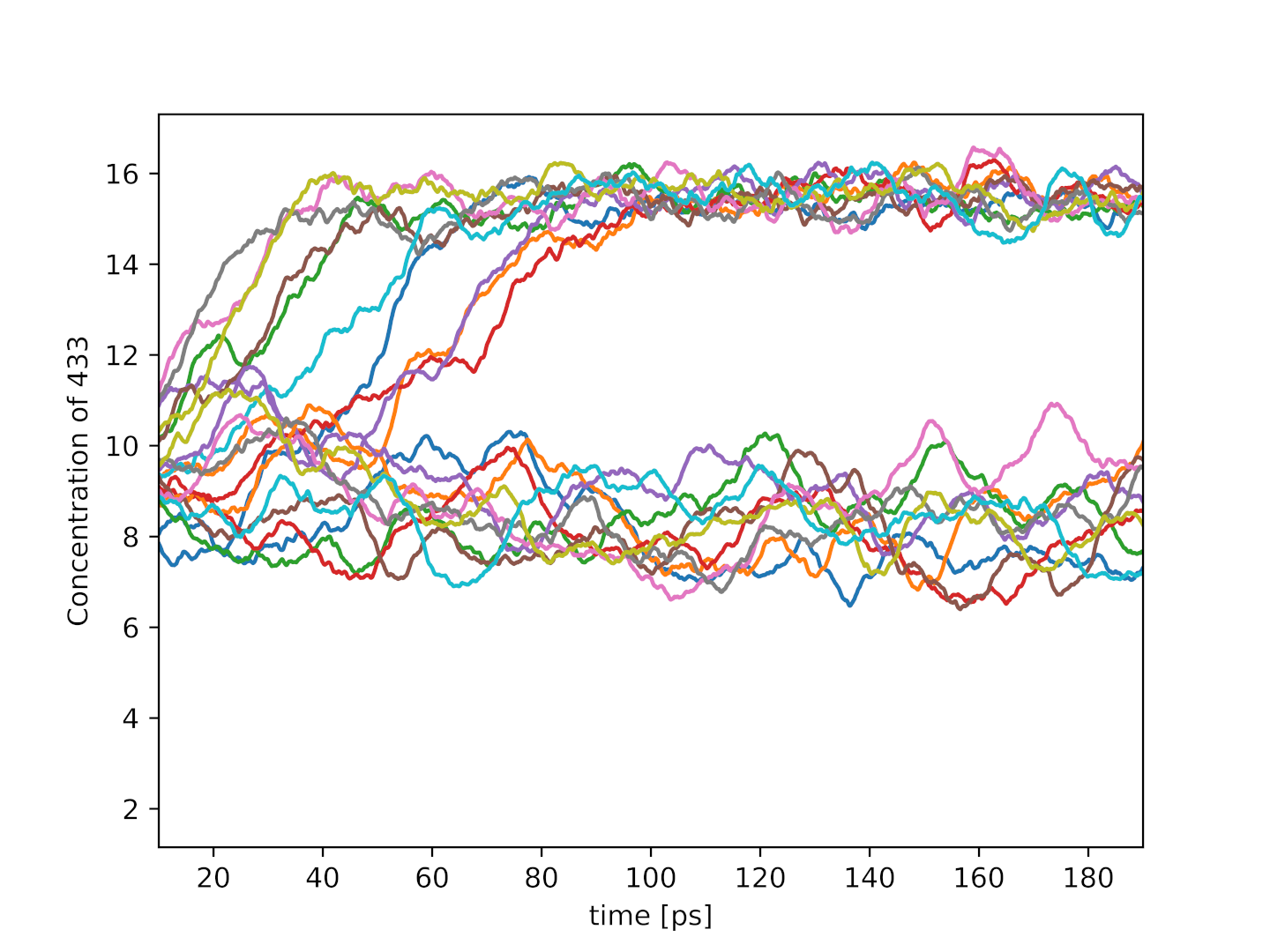}
        \caption{433}
    \end{subfigure}
    \caption{Configurations triplets related to the presence of perfect (555) and distorted (544 and 433) icosahedra. These distorted triplets are also related to liquid and
liquid-like phases~\cite{Xiao2006b,UrrutiaBunuelos2016}.}
    \label{fig:CNA-Others}
\end{figure*}

Additionally, Figure~\ref{fig:CNA-Others} displays another set of triplets. These correspond to the 555, 544 and 433 triplets, widely observed during simulations. The 555 is characteristic
of a perfect icosahedral order observed in the structure, meanwhile the 544 and 433 are typical distorted icosahedral configurations~\cite{ValenciaBalvin2010,Xiao2006b}. These structures are
present on both cases, critical superheated solid and liquid state. Despite the above, a clear increase of perfect icosahedra is observed in the liquid state, raising from $\sim 10$\% to
values closer to $\sim 30$\%. Interestingly, the presence of distorted icosahedra remains almost constant on both phases, with a percentage of presence in the structure of $\sim 14$\%.
These structures are frequent in liquid and liquid-like phases~\cite{Xiao2006b,UrrutiaBunuelos2016}. The continuous presence of liquid-like short-range order (distorted 544 icosahedra) in the metastable solid might hint at a precursor of the crystal collapse.

\section{\label{sec:Conclusions}Conclusions}

We have found that, after a single primary knock-on atom (PKA) event on a bcc W crystal, is possible to drive it into a critical superheated solid phase, which is metastable
and decays after about 200 ps. The existence of the metastable state is verified by the melting waiting times, whose statistical distribution is shown to be similar to that of
the superheated state observed in homogeneous melting via the Z-method.

Characterization of the metastable structure, and a detailed description of both states were presented using a selected subset of simulations, which suggests the presence of
particular triplets in the CNA analysis that might serve as precursors of melting.

Our results provide an additional mechanism to study the microscopic dynamics leading to melting from an ideal crystal.

\section{\label{sec:Acknowledgements}Acknowledgements}

This work was supported by FONDECYT Iniciaci\'on grant 11150279 and UNAB DI-1350-16/R (C.L.). S.D. acknowledges support from Anillo ACT-172101
and FONDECYT 1171127 grants. J.P. acknowledges support from UNAB DI-15-17/RG. Computational work was supported by the supercomputing infrastructures of the NLHPC (ECM-02), and FENIX
(UNAB).

\section{\label{sec:DataAvail}Data availability}
 The raw/processed data required to reproduce these findings cannot be shared at this time due to technical or time limitations.


\providecommand{\noopsort}[1]{}\providecommand{\singleletter}[1]{#1}%

\end{document}